\documentstyle[12pt]{article}
\newcommand{\beq}{\begin{equation}}
\newcommand{\eeq}{\end{equation}}
\newcommand{\cD}{{\cal D}}
\newcommand{\cB}{{\cal B}}
\newcommand{\ob}{\vert_{\partial M}=0}
\begin{document}
\title{Gauge-invariance in one-loop quantum cosmology
\thanks{A talk presented at 6th Seminar on Quantum Gravity,
Moscow, June 1995}}
\author{Dmitri V.Vassilevich}
\date{ }
\maketitle
\centerline{\it Department of Theoretical Physics, St.Petersburg
University,}
\centerline{\it 198904 St.Petersburg, Russia \/}

    \begin{abstract}
We study the problem of gauge-invariance and gauge-dependence
in one-loop quantum cosmology. We formulate some requirements
which should be satisfied by boundary conditions in order to
give gauge-independent path integral. The case of QED is
studied in some detail. We outline difficulties in
gauge-invariant quantization of gravitational field in a
bounded region.

    \end{abstract}

\section{Introduction}

    According to the Hartle--Hawking prescription [1], the
wave function of the Universe is given by the Euclidean path
integral over all four metrics with a given boundary
three-geometry and all configurations of matter fields with
some boundary conditions (for a review, see recent monograph
[2]). Thus, at least at one-loop approximation the quantum
cosmology is equivalent to the theory of path integration in
a bounded region of the Euclidean space.

    Past years, many contradicting results were reported
for the scaling coefficient $\zeta (0)$ calculated in
different approaches. However, during last year most of
these contradictions were removed. It was demonstrated [3]
that the direct eigenvalue method and analytical expressions
in terms of geometric quantities give equivalent results
if one takes into account corrections [4] to the formulae of
[5]. It was shown [6] that disagreement between covariant
path integral and that over physical degrees of freedom
is due to the use non-covariant measure in the latter one.
Complete agreement is restored if one uses covariant path
integral measure.

    At the same time, some recent computations [7,8,9]
indicate that the one-loop amplitudes on the Euclidean
four-disk are gauge-dependent even in a framework of
the same direct eigenvalue method. This problem is studied
in the present work. We demonstrate, that to obtain
gauge-independent path integral one should observe
some consistency conditions, which are formulated
in the next section. In the section 3 we show that
these consistency conditions are violated whenever
gauge-dependence was observed in QED. We also comment
on invariant gauge conditions for gravitational perturbations.

\section{Gauge-invariant boundary conditions}

  A gauge-invariant formulation of the path integral
for gauge theories was suggested a long ago by Faddeev and Popov
[10]. Consider a standard proof [11] of gauge-independence of
path integral paying special attention to specific features
of manifolds with boundaries.

    Let $\Phi$ be a gauge field. Consider the path integral in
some gauge $\chi$ with a gauge-fixing term added to classical
action $S(\Phi )$:
\beq
Z(\alpha , \chi )=\int \cD \Phi J(\chi )
\exp (-S(\Phi ) -\frac 1{2\alpha} \chi^2)
\label{eq:gf}
\eeq
Integration over space-time coordinates is assumed. $J(\chi )$
is the Faddeev-Popov determinant corresponding the gauge
condition $\chi $. It is well known that the path integral
(\ref{eq:gf}) can be obtained from another path integral
\beq
Z(\chi_a ) =\int \cD \Phi J(\chi ) \delta (\chi -a)
\exp (-S(\Phi )) \label{eq:ca}
\eeq
after averaging over $a$ with the weight $\exp
(-\frac 1{2\alpha} a^2)$. There is nothing specific at this
point with respect to manifolds with boundaries. One can
self-consistently define boundary conditions for $a(x)$
through boundary conditions for $\chi (\Phi )$. Hence,
it is enough to study gauge-independence of the path
integral
\beq
Z(\chi ) =\int \cD \Phi J(\chi ) \delta (\chi )
\exp (-S(\Phi )) \ . \label{eq:c}
\eeq
The equivalence of two path integrals,
$Z(\chi_1 )$ and $Z(\chi_2 )$ can be established by using the
Faddeev-Popov trick. One should use twice the following
representation of unity
\beq
1=\int \cD \xi J(\chi ) \delta (\chi (\Phi +\delta_\xi \Phi )),
\label{eq:un}
\eeq
where $\delta_\xi \Phi$ are (linearized) gauge transformations
of the field $\Phi$. One should insert (\ref{eq:un}) with
$\chi =\chi_2$ in the integrand of $Z(\chi_1)$, change
integration variables to $\Phi -\delta_\xi \Phi$, and use
again eq. (\ref{eq:un}) with $\chi =\chi_1$. This procedure
can be done successfully if the two gauges $\chi_1$ and
$\chi_2$ satisfy the following requirements.

    (i) {\it Gauge-invariance of the boundary conditions}.
Let
\beq
\cB \Phi \ob \label{eq:bp}
\eeq
be a boundary condition for the fields $\Phi$ with some
boundary operator $\cB$, which describes Dirichlet, Neumann
or mixed boundary conditions. There should exist boundary
conditions
\beq
{\cB}_\xi \xi \ob \label{eq:bx}
\eeq
for gauge transformation parameters $\xi$ such
that
\beq
\cB \delta_\xi \Phi \ob \ . \label{eq:inv}
\eeq

    The eq. (\ref{eq:inv}) means that gauge transformations
map the functional space defined by eq. (\ref{eq:bp}) onto
itself for some boundary conditions (\ref{eq:bx}) imposed
on gauge parameter $\xi$.

    (ii) {\it Simultaneous admissibility of $\chi_1$ and
$\chi_2$}. We call a gauge condition $\chi$ admissible
if for given gauge-invariant boundary conditions
(\ref{eq:bp}), (\ref{eq:bx}) the equation
\beq
\chi (\Phi +\delta_\xi \Phi )=0 \label{eq:adm}
\eeq
has unique solution{\footnote{Note that we consider
only linearized gauge transformations thus
avoiding the question of Gribov ambiguities.
This restriction is correct at least at one-loop
approximation}}
 $\xi$ for every $\Phi$. The gauges $\chi_1$ and
$\chi_2$ should be admissible for the same
boundary operators $\cB$ and $\cB_\xi$.

    Using the Faddeev-Popov trick one can easily
demonstrate that $Z(\chi_1 )$$=Z(\chi_2 )$ with
the integration regions defined by (\ref{eq:bp})
with the same boundary operator $\cB$ provided
the conditions (i) and (ii) are satisfied.

\section{QED in a bounded region}

    Consider a simple example of QED on four-dimensional
unit disk. The metric has the form
\beq
ds^2=dx_0^2+x_0^2 d\Omega^2 \ ,\label{eq:met}
\eeq
where $d\Omega^2$ is the line element on unit three-sphere.

    Let us define gauge-invariant boundary conditions for
electromagnetic vector potential $A_\mu$. Gauge transformations
of the $A_\mu$ are
\beq
\delta A_\mu =\partial_\mu \omega (x) \ . \label{eq:tra}
\eeq
Suppose, that the boundary operator $\cB_\omega$ is local and
$SO(3)$-invariant. This means, that $\omega$ satisfies either
Dirichlet, $\cB_\omega =1$, or Neumann boundary conditions,
$\cB_\omega =\partial_0 +C$, with some constant $C$. It is
easy to see, that for the former case the only gauge invariant
local boundary conditions for $A_\mu$ are relative or magnetic
boundary conditions:
\beq
A_i \ob ,\quad (\partial_0 +3)A_0 \ob \label{eq:rel}
\eeq
$i=1,2,3.$ In the latter case it is possible to define
gauge-invariant local boundary conditions only for $C=0$.
These are the so-called absolute or electric boundary
conditions:
\beq
\partial_0 A_i \ob ,\quad A_0 \ob \label{eq:abs}
\eeq
The eqs. (\ref{eq:rel}) and (\ref{eq:abs}) give the only
local gauge-invariant boundary conditions. Of course, if
we abandon the locality requirement, we obtain more
boundary conditions satisfying (i).

    The conditions (\ref{eq:rel}) and (\ref{eq:abs}) have
one extra property, which makes them preferable. The
quadratic form of the action is represented by a hermitian
second order differential operator.

    Let us now analyze gauge conditions satisfying the
requirement (ii). As a reference gauge $\chi_1$ let us
choose the Lorentz gauge
\beq
\nabla^\mu A_\mu =\chi_1=0 \ .\label{eq:lor}
\eeq
This gauge is admissible for both absolute and relative
boundary conditions. Let check whether some popular
gauges can be chosen as $\chi_2$.

{\it The Coulomb gauge}. In our coordinate system the
Coulomb gauge condition takes the form
\beq
\ ^{(3)}\nabla^i A_i =\chi_c=0 \label{eq:col}
\eeq
Here $\ ^{(3)}\nabla_i$ denotes the covariant
derivative with respect to three-metric.
It is easy to see, that the condition (\ref{eq:col})
does not fix the gauge freedom completely. The gauge
transformations with $x_i$-independent parameter,
$\partial_i \omega =0$, $i=1,2,3$, remain unfixed.
The space of such transformations is spanned by a
one-parameter family of eigenfunctions of the scalar
Laplace operator, $(x^0)^{-1}J_1(x^0\lambda )$, where
$J_1$ is the Bessel function; the eigenvalues $\lambda$
are defined by one of the conditions, $J_1(\lambda )=0$
or $(\partial -1)J_1(\lambda )=0$, depending on the
boundary operator $\cB_\omega$. This means that an
additional gauge fixing condition is needed. One can
use e.g.
\beq
\tilde \chi =<A_0>=0, \label{eq:add}
\eeq
where $<A_0>$ denotes average value of $A_0$ on a spatial
slices. The condition (\ref{eq:add}), however, can not
be represented in a convenient way as gauge-fixing term
in the action. In the view of the above consideration
the disagreement between Lorentz and Coulomb gauges on
a disk reported recently [8] looks quite natural

    {\it Temporal gauge}. In our coordinate system the
temporal gauge $A_0=0$ in fact coincides with the Fock
radial gauge [12]. At first glance, this gauge condition
does not fix the gauge freedom corresponding to gauge
parameter depending on spatial coordinates only.
However, looking at scalar harmonics on unit disk,
which have the form $(x^0)^{-1}J_{l+1}(\lambda x^0)Y_l(x_i)$,
we see, that all harmonics with non-trivial dependence
on spatial coordinates, $l \ge 1$, have zero in the
origin of the coordinate system and thus can not be
$x^0$-independent. This means that on a disk the temporal
gauge fixes the gauge freedom completely. It can be seen,
that this gauge is admissible in the sense of (ii).

    For the boundary conditions (\ref{eq:rel}) and
(\ref{eq:abs}) the Hodge--de Rham decomposition
\beq
A_\mu =A_\mu^\perp +\partial_\mu \omega \quad
\nabla^\mu A_\mu^\perp =0 \label{eq:hod}
\eeq
is orthogonal with respect to ordinary inner product
in the space of vector fields without surface terms.
The Jacobian factor of the change of variables
$\{ A\} \to \{ A^\perp ,\omega \}$ is just $J^{\frac 12}$,
where $J$ is the ghost determinant in the Lorentz gauge.
For any admissible gauge condition $\chi$ one can
express a solution of equation $\chi (A)=0$ in the form
$A=A^\perp +\partial \omega (A^\perp )$. Thus $A^\perp$
can be used as coordinates on the space of solutions of
a gauge condition $\chi$. In this coordinates the
equivalence between Lorentz path integral and that in
the gauge $\chi$ becomes evident.

    {\it The Esposito gauge}. The Esposito gauge condition
[2,7] reads
\beq
\chi_{\rm Esp}=\partial_0 A_0 +^{(3)}\nabla^i A_i=
\partial^\mu A_\mu -\frac 3{x^0} A_0 \ .\label{eq:esp}
\eeq
Let us decompose spatial components $A_i$ in longitudinal
and transversal parts:
\beq
A_i=A_i^T+\partial_i s \ , \quad \ ^{(3)}\nabla^iA_i^T=0
\label{eq:ho3}
\eeq
The eq. (\ref{eq:esp}) gives
\beq
\partial_0 A_0 +^{(3)}\Delta s =0 \label{eq:ess}
\eeq
where $\ ^{(3)}\Delta$ is the Laplace operator with
respect to three-metric. Eq. (\ref{eq:ess}) makes it
possible to express $s$ in terms of $A_0$ and thus
eliminate one scalar degree of freedom, as any gauge
condition should do. Consider now the condition
(\ref{eq:ess}) on the boundary. Let $A_\mu$ satisfy
the relative boundary conditions (\ref{eq:rel}).
We have
\beq
\chi_{\rm Esp}(A_\mu ) \vert_{\partial M}=
-3A_0 \ob \label{eq:esb}
\eeq
This means that $A_0$ should in the same time satisfy
Dirichlet and Newmann boundary conditions. Thus one
more degree of freedom is excluded, and the Esposito
gauge is incompatible with relative boundary conditions.
The same is also true for absolute boundary conditions.
This explains discrepancies [9] between Lorentz and Esposito
gauges on a disk.

    \section{Quantum gravity}

    The problem of formulation of gauge-invariant boundary
condition for gravitational perturbations in much more
complicated than that for electromagnetic filed. One
can demonstrate [13,14], that there are no gauge-invariant boundary
conditions for quantum gravity, which are local for both
graviton and ghost perturbations if boundary is
not totally geodesic. Such boundary conditions
can be formulated for quantum gravity with dynamical torsion
in two dimensions [13]. However, it is not clear, whether
this result can be extended to higher dimensions. Though
the locality requirement seems to be technical, local
boundary conditions almost automatically lead to
self-adjointness of the quadratic form of the action.
For example, the Luckock-Moss-Poletti boundary conditions [15]
do really lead to self-adjoint Laplace operator. Unfortunately,
these boundary conditions are only partially invariant,
in the agreement with the above statement.

    At present, the non-local boundary conditions suggested
by Barvinsky [16] are the best choice. These boundary conditions
are gauge invariant. Recently, manifest computations on a disk
were performed [14]  and a new class of non-local boundary
conditions was suggested [17]. However, self-adjointness of
the quadratic form of the action has not been proved.

    One can formulate most general gauge invariant
boundary conditions for graviton fluctuations giving
self-adjoint action at least on a disk [18]. Unfortunately,
these boundary conditions have a very complicated form
and are hardly suitable for actual computations.
Probably, a more careful analysis of classical boundary
problem is needed in order to formulate basic properties
of quantum gravity in a bounded region.

    \section{Discussion}

    To the best of our knowledge, in all cases when
gauge-dependence of on-shell amplitudes in one-loop
quantum cosmology was observed, at least one of the
requirement of Sec. 2 is violated. However, it was
demonstrated that for QED in a region between two
concentric spheres the $\zeta (0)$ is gauge
independent even if (ii) is violated [7-9]. Two
explanations to this fact are possible. First, that
the $\zeta (0)$, being odd function of the orientation
of normal vector on a boundary, is not sensitive to
gauge non-invariant part of the path integral.
Second, that since in this region a smooth 3+1
split is possible actual invariance of the path
integral is higher than predicted for general
case. A simple test which could help to choose
between these two explanation may be a computation
of more terms of the heat kernel expansion and/or
computation of $\zeta (0)$ for even-dimensional
boundary.


    The author is grateful to Andrei Barvinsky, Giampiero
Esposito and Alexander Kamenshchik for discussions and
correspondence. This work was supported by the Russian
Foundation for Fundamental Studies, grant 93-02-14378.

    \section*{References}
1. J.B. Hartle and S.W. Hawking, Phys. Rev. {\bf D28}
(1983) 2960.
\newline
2. G. Esposito, Quantum Gravity, Quantum Cosmology and
Lorentzian Geometries, Springer, Berlin, 1992
\newline
3. I. Moss and S. Poletti, Phys. Lett. {\bf B333} (1994)
326.
\newline
4. D.V.Vassilevich, Vector fields on a disk with mixed
boundary conditions, SPbU-IP-94-6, gr-qc/9405052. to
appear in J. Math. Phys.
\newline
5. T.P. Branson and P.B. Gilkey, Commun. Part. Diff. Eqs {\bf
15} (1990) 245.
\newline
6. D.V. Vassilevich, QED on a curved background and on
manifolds with boundaries: unitarity vs covariance, IC/94/359,
gr-qc/9411036, to appear in Phys. Rev. D.
\newline
7. G. Esposito, A.Yu. Kamenshchik, I.V. Mishakov and
G.Pollifrone, Class. Quantum Grav. {\bf 11} (1994) 2939.
\newline
8. G. Esposito and A.Yu.Kamenshchik, Phys. Lett. {\bf B336}
(1994) 324.
\newline
9. G. Esposito, A.Yu. Kamenshchik, I.V. Mishakov and
G.Pollifrone, Relativistic gauge conditions in quantum
cosmology, gr-qc/9504007.
\newline
10.
L.D. Faddeev and V.N. Popov, Phys. Lett. {\bf 25B} (1967) 29.
\newline
11. L.D. Faddeev and A.A. Slavnov, Gauge Fields: Introduction to
Quantum theory, Benjamin/Cummings, 1980.
\newline
12. V.A.Fock, Sov. Phys. {\bf 12} (1937) 404.
\newline
13. D.V. Vassilevich, On gauge-invariant boundary condition for
2d gravity with dynamical torsion, Preprint TUW 95-06,
hep-th/9504011.
\newline
14. G. Esposito, A.Yu. Kamenshchik, I.V. Mishakov and
G.Pollifrone, One-loop amplitudes in Euclidean quantum gravity,
DSF preprint 95/16.
\newline
15. I. Moss and S.Poletti, Nucl. Phys. {\bf B245} (1990) 355.

H. Luckock, J. Math. Phys. 32 (1991) 1755.
\newline
16. A. Barvinsky, Phys. Lett. {\bf B195} (1987) 344.
\newline
17. G. Esposito, A.Yu. Kamenshchik, I.V. Mishakov and
G.Pollifrone, Non-local boundary conditions in Euclidean
quantum gravity.
\newline
18. V.N. Marachevsky and D.V.Vassilevich, work in preparation.

\end{document}